\documentclass[10pt,a4paper,twocolumn,showpacs,showkeys,prd,balancelastpage]{revtex4}
\usepackage{amsfonts}
\usepackage{amsmath}
\usepackage{amssymb}
\usepackage{graphicx}

\begin{document}

\title[ ]{Proposed split-causality test of the relativity principle}
\author{George Jaroszkiewicz}
\affiliation{School of Mathematical Sciences, University of Nottingham, Nottingham, UK}
\keywords{relativity principle, quantum causality, quantum horizon}
\pacs{03.30.+p, 03.65.-w}

\begin{abstract}
We propose a test of the principle of relativity, involving quantum signals
between two inertial frames. If the principle is upheld, classical causality
will appear to be split in a dramatic and emphatic way. We discuss the
existence of quantum horizons, which are barriers to the transmission of any
form of quantum information. These must occur in any finite time,
inter-frame experiment if quantum causality holds. We conclude with some
comments on such experiments involving entangled states.
\end{abstract}

\date{\today}
\maketitle

\section{Introduction}

The principle of relativity in its strong form states that the laws of
physics are the same in all standard inertial frames, if gravitational
effects are excluded. It was used by Einstein to derive the Lorentz
transformation
\begin{equation}
t^{\prime }=\gamma \left( t-vx/c^{2}\right) ,\ x^{\prime }=\gamma \left(
x-vt\right) ,\ y^{\prime }=y,\ z^{\prime }=z,  \label{111}
\end{equation}%
$\gamma =1/\sqrt{1-v^{2}/c^{2}}$, between two standard inertial frames $%
\mathcal{F}$, $\mathcal{F}^{\prime }$ moving apart with relative speed $v$
along the $x$-direction \cite{EINSTEIN-1905}. A notable feature is the loss
of absolute simultaneity: a hyperplane of simultaneity $t^{\prime }=const$
in $\mathcal{F}^{\prime }$ is not a hyperplane of simultaneity in $\mathcal{F%
}$. We shall investigate the consequences of adding quantum physics to this
feature of relativity.

In contrast to most approaches to relativity, we shall focus on very
specific hyperplanes of simultaneity, one in each frame. For convenience, we
shall ignore the $y$ and $z$ coordinates; although they are physically
relevant, they do not feature in the essential points of the discussion. We
shall consider an experiment where a signal is sent from $\mathcal{F}$ at
time $t=0$ and observed in frame $\mathcal{F}^{\prime }$ at time $T^{\prime
} $, as measured in its frame. Figure $1$ shows the essential details.

\begin{figure}[t]
\centerline{\includegraphics[width=3.0in]{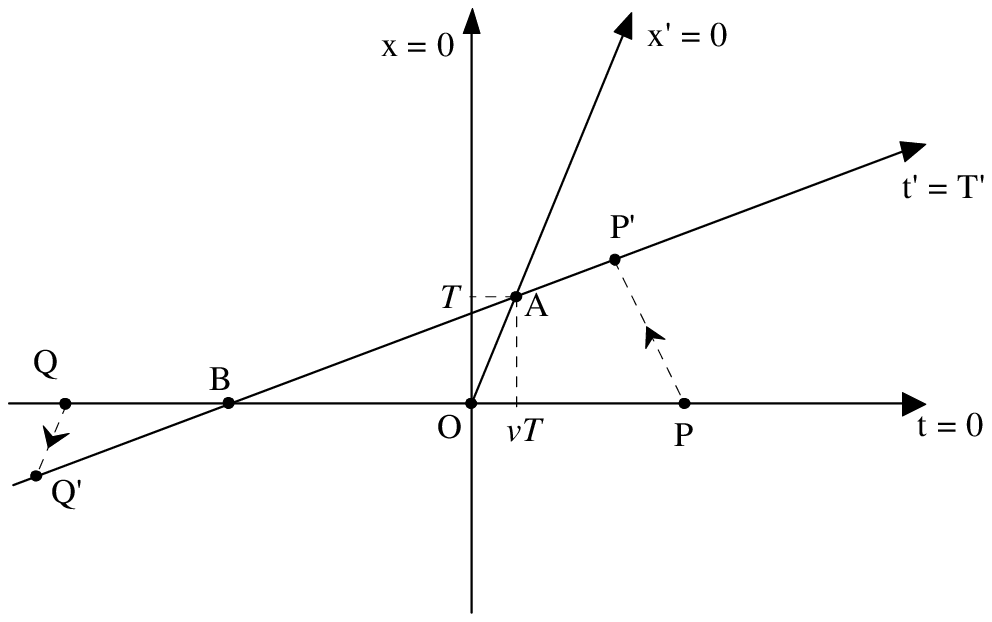}}
\caption{$B$ is the quantum horizon. }
\end{figure}

Event $O$ is the common origin of spacetime coordinates, events $P$ and $Q$
are simultaneous in $\mathcal{F}$ at time $t=0$ whilst events $P^{\prime }$,
$Q^{\prime }$ are simultaneous in $\mathcal{F}^{\prime }$ at time $t^{\prime
}=T^{\prime }$. Our convention is that an event $P$ has coordinates $%
(t_{P},x_{p})$ in $\mathcal{F}$, coordinates $\left[ t_{P}^{\prime
},x_{P}^{\prime }\right] $ in $\mathcal{F}^{\prime }$ and we write $P\sim
(t_{P},x_{P})\sim \lbrack t_{P}^{\prime },x_{p}^{\prime }]$. A critical
feature is event $B$, where the lines of simultaneity $t=0$ and $t^{\prime
}=T^{\prime }$ intersect. Assuming $v$ is positive, then for $B$ we have
\begin{equation}
B\sim (0,-\frac{c^{2}T^{\prime }}{\gamma v})\sim \lbrack T^{\prime },-\frac{%
c^{2}T^{\prime }}{v}].
\end{equation}

Recall that in de Broglie wave mechanics, the speed $w$ of a pilot wave
associated with a physical particle moving with subluminal speed $v$
satisfies the relation $vw=c^{2}$. This suggests that such a pilot wave
cannot be used to convey physical signals, because it travels at
superluminal speed. The position of event $B$ appears to observers in $%
\mathcal{F}^{\prime }$ to be the wave front at time $T^{\prime }$ of such a
pilot wave associated with frame $\mathcal{F}$, if it were sent out from $O$
in the same direction. Clearly, for large $T^{\prime }$ and small $|v|$, $B$
will be at relatively large distance from the origin of spatial coordinates $%
A$ in frame $\mathcal{F}^{\prime }$, at the time $t^{\prime }=T^{\prime
}=T/\gamma $ when the signal from $\mathcal{F}$ is observed in $\mathcal{F}%
^{\prime }$ during the experiment.

In standard discussions of special relativity, event $B$ is generally
ignored, as it appears to be far removed from events $P$ and $P^{\prime }$
involved in the signalling experiment. For this experiment, we imagine that
a quantum state has been prepared by apparatus $\mathcal{A}_{P},$ at rest in
frame $\mathcal{F}$, and a contingent quantum outcome subsequently detected
by apparatus $\mathcal{A}_{P^{\prime }}^{\prime }$, at rest in frame $%
\mathcal{F}^{\prime }$.

The critical word here is \textquotedblleft \emph{subsequently%
\textquotedblright . }Quantum physics, \textbf{as it is performed in real
laboratories}, can discuss only the possibility of quantum information
travelling forwards in time. Both signal emitter and signal detector in any
quantum experiment must agree that the former acts before the latter.
Otherwise, the physical significance of the Born probability rule would be
completely undermined. In quantum theory and in the real world, we do not
know the outcome of an experiment before it is performed. We shall call the
requirement that $P$ is earlier than $P^{\prime }$ in both frames of
reference \emph{quantum causality.}

From Figure $1$, it is clear that there is no problem with quantum causality
as far as events $P$ and $P^{\prime }$ are concerned. But consider events $Q$
and $Q^{\prime }$ on the other side of $B$. If quantum causality is valid,
then signals prepared at $Q$ cannot be received by $Q^{\prime }$. In
essence, event $B$ acts a barrier to quantum causality, and on this account
we shall refer to $B$ as a \emph{quantum horizon. }

Ordinarily, such quantum horizons are ignored in conventional physics,
because under most circumstances, $B$ appears to be very far from events
such as $P$ and $P^{\prime }$. In experiments looking at such aspects of
quantum information, speeds in excess of $10^{5}$ $c$ have been reported
\cite{SCARANI-2000}. In practice, high energy particle theory conventionally
takes the scattering limit $T^{\prime }$ $\rightarrow $ $\infty $, $v=0$ in
the calculation of Lorentz covariant matrix elements. Finite-time processes
and inter-frame experiments of the sort considered by us here are generally
avoided, because it is assumed there is no significant novel physics
involved. An important factor in this is that the scattering limit makes
calculations such as those arising from Feynman diagrams take on relatively
standard forms and does not involve issues to do with a quantum horizon,
which is at spatial infinity under those circumstances. Such a
simplification does not happen for finite-time and inter-frame processes.

We now consider the implications of the relativity principle and ask the
following question: if according to the relativity principle frames $%
\mathcal{F}$ and $\mathcal{F}^{\prime }$ are \textquotedblleft just as good
as each other\textquotedblright , why does the quantum horizon $B$ appear to
distinguish between the two?

A little though soon resolves the question. If the relativity principle is
valid, then there must be a symmetry between the two frames. There is no
doubt that a quantum signal can be prepared at $P$ and received at $%
P^{\prime }$, if $P^{\prime }$ is in or on the forwards lightcone with
vertex $P.$ Quantum causality rules out the transmission of a quantum signal
from $P^{\prime }$ to $P$, and the transmission of a signal from $Q$ to $%
Q^{\prime }$. But nothing currently known in physics forbids the possibility
of a physical signal being sent from $Q^{\prime }$ to $Q$, if $Q$ is in the
forwards lightcone with vertex $Q^{\prime }$. Indeed, symmetry demands such
a possibility. This is the essence of the split causality experiment
proposed here.

\section{Proposed experiment}

Based on the above considerations, we propose the following experimental
test of the principle of special relativity. It will undoubtedly be
difficult to perform, but would test the principle of relativity in a
spectacular and convincing way.

We envisage the use of four spacecraft $P,Q$, $P^{\prime }$ and $Q^{\prime }$%
, sufficiently far from gravitating bodies to justify the use of the special
relativistic transformation rules (\ref{111}). $P$ and $Q$ are in the same
rest frame $\mathcal{F}$ and situated at some distance from each other. By
prior signalling arrangement, clocks on $P$ and $Q$ craft have been
synchronized. Likewise, $P^{\prime }$ and \thinspace $Q^{\prime }$ are in
their own rest frame $\mathcal{F}^{\prime }$ and all their clocks have been
synchronized.

With reference to Figure $1$, spacetime homogeneity means that we may always
transfer the origin of spacetime coordinates in both frames $\mathcal{F}$, $%
\mathcal{F}^{\prime }$ to the quantum horizon $B$. This means that the
hyperplanes of simultaneity involved in the experiment are now at times $t=0$
in $\mathcal{F}$ and $t^{\prime }=0$ in $\mathcal{F}^{\prime }$, as shown in
Figure $2$.

The experiment consists of $P$ sending a brief light pulse signal towards $%
P^{\prime }$ at time $t=0$, whilst simultaneously in $\mathcal{F}$, $Q$
opens a detector in order to receive light from $Q^{\prime }$ for a similar
brief period. In addition, the same protocol is carried out in frame $%
\mathcal{F}^{\prime }$ at time $t^{\prime }=0$: $Q^{\prime }$ sends a brief
light pulse towards $Q$ whilst simultaneously in $\mathcal{F}^{\prime }$, $%
P^{\prime }$ opens a detector to receive a signal from $P$. The whole
experiment is illustrated in Figure $2$.

\begin{figure}[t]
\centerline{\includegraphics[width=3.0in]{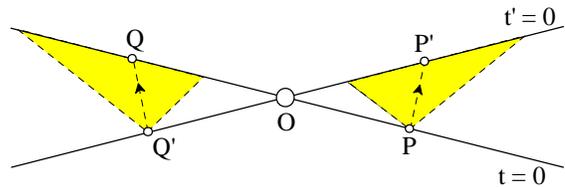}}
\caption{$O$ is the quantum horizon. Shaded regions are forwards lightcones.
Signals shown with subluminal transmission speeds, which do not alter the
overall conclusions.}
\end{figure}

After the signals have been sent and received, observers from all spacecraft
can meet at leisure and compare results. If it turns out that $P$ sent a
signal at the same time $t=0$ that $Q$ received a signal, and that $%
P^{\prime }$ received a signal at the same time $t^{\prime }=0$ that $%
Q^{\prime }$ sent a signal, then the principle of relativity would be upheld
in a most convincing way. Quantum causality would have been respected but
classical causality would appear to be \textquotedblleft
split\textquotedblright\ in a most remarkable and counterintuitive fashion.
On the other hand, if no such result was ever detected despite repeated
attempts, this would rule out the principle of special relativity and
undermine the whole of conventional physics. Alternatively, quantum
causality would have to be reviewed, with equally disastrous implications
for modern physics.

Although the result may appear a forgone conclusion in favour of the
relativity principle, complacency here would be unjustifiable. An actual
experiment involving relatively moving frames has to be involved. It would
not do to simulate the experiment using a single frame of reference and
invoking scattering processes involving high energy particles, unless the
issues of the meaning of timing of virtual particle processes which would
inevitably arise were adequately resolved.

Such an inter-frame experiment would not need to be performed more than once
to establish the principle of relativity in the most spectacular way; a
single splitting of causality would suffice to vindicate both the Lorentz
transformation rule (\ref{111}) and our insistence on the need for the
quantum causality rule.

The difficulties in this experiment are of course technological. First, it
would be expensive though not impossible to arrange four spacecraft in such
a configuration. Current resources rule out such a possibility, but we
envisage that with the present increasing interest in the colonization of
the Moon and subsequent journeys to Mars, space travel will develop into a
more routine activity, with an increased availability of vehicles for such
an experiment.

Some economy could be found by using the Moon and some other body, such as
the Earth or a suitable asteroid, to locate say events $P$ and $Q$. A
significant residual problem would be the need to boost the two spacecraft
representing events $P^{\prime }$ and $Q^{\prime }$ to sufficiently high
speed $|v|$ to allow unambiguous effects to be observed. An even greater
economy could be made by using four small unmanned but self-propelled
probes, two of which were programmed to emit signals whilst the other two
were programmed to receive them. Such an experiment seems within the
capability of present-day technology. We note that the signals involved need
not be light signals. Any form of communication with subluminal propagation
speeds in the proposed set-up should demonstrate classical split-causality,
if the principle of relativity holds.

\section{Entangled states and quantum horizons}

The above scenario involves classical signalling processes. An even more
interesting situation arises when quantum entangled states are involved, as
this touches upon the debate concerning information loss in black hole
physics. Consider the experiment illustrated by Figure $3$, where a spin
zero positronium state is prepared at event $S$. The forwards
lightcone with vertex $S$ contains events $P,Q$, $P^{\prime }$ and $%
Q^{\prime }$, so that in principle each event could detect component
particles of the prepared state. We shall imagine restricting $P$ and $%
P^{\prime }$ to the observation of electron spin orientation only, via
Stern-Gerlach apparatus, whilst $Q$ and $Q^{\prime }$ do the same for
positrons.

\begin{figure}[t]
\centerline{\includegraphics[width=3.0in]{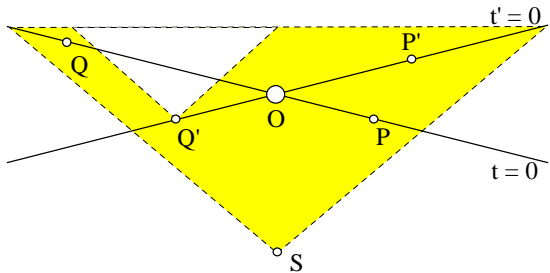}}
\caption{A positronium state is prepared as $S$. Events $P,Q$, $P^{\prime }$
and $Q^{\prime }$ are well with the forwards lightcone centred on $S$. }
\end{figure}

Depending on the choices made, a number of different observation protocols
could be implemented, but not necessarily all in the same run (we assume the
overall experiment is repeatable as often as required to collect adequate
statistics). It is useful here to assign a different Hilbert space to each
such observation protocol: $\mathcal{H}\left( P\right) $ is the Hilbert
space used to describe the potential outcome of an electron spin test at $P$%
, $\mathcal{H}\left( P\cup Q\right) =\mathcal{H}\left( P\right) \otimes
\mathcal{H}\left( Q\right) $ is the Hilbert space used to describe a
simultaneous-in-$\mathcal{F}$ observation of an electron at $P$ and a
positron at $Q$, and so on.

Provided $Q^{\prime }$ was instructed in advance not to perform its test,
then $P$ and $Q$ could perform their tests, describing the state prepared by
$S$ in terms of an entangled state in $\mathcal{H(}P\cup Q\mathcal{)}$ of
the form $u_{P}\otimes d_{Q}-d_{P}\otimes u_{Q}$, where $u_{P}$ represents a
spin-up outcome at $P$, $d_{Q}$ represents a spin-down outcome at $Q$, and
so on. Likewise, provided $P$ was instructed in advance not to perform its
test, then $P^{\prime }$ and $Q^{\prime }$ could perform their tests,
describing the prepared state in term of the entangled state $u_{P^{\prime
}}^{\prime }\otimes d_{Q^{\prime }}^{\prime }-d_{P^{\prime }}^{\prime
}\otimes u_{Q^{\prime }}^{\prime }$, an element in $\mathcal{H}(P^{\prime
}\cup Q^{\prime })=\mathcal{H(}P^{\prime })\otimes \mathcal{H}(Q^{\prime })$%
. Conservation of electric charge rules out any possibility of outcomes
described via $\mathcal{H}(P)\otimes \mathcal{H}(P^{\prime })$ or $\mathcal{H%
}(Q)\otimes \mathcal{H}(Q^{\prime })$. However, in principle, it should be
possible to involve outcomes described in $\mathcal{H}(P)\otimes \mathcal{H}%
(Q^{\prime })$ or $\mathcal{H(}P^{\prime })\mathcal{\otimes H(}Q\mathcal{)}$.

An interesting possibility arises with an experiment described initially in $%
\mathcal{H}\left( P\right) \otimes \mathcal{H}\left( Q\right) $. In such an
experiment, $Q$ would always observe a positron whenever $P$ observed an
electron. Now consider the addition of a choice at $Q^{\prime }$ to test
positron spin. Any observation at $Q^{\prime }$ would take place before $Q$,
according to frames $\mathcal{F}$ and $\mathcal{F}^{\prime }$, and therefore
quantum causality would apply. The detection of a positron at $Q^{\prime }$
would destroy the possibility of a detection of a positron at $Q$, because
of charge conservation. This would hold even if $Q$ was outside the forwards
lightcone centred on $Q^{\prime }$, as shown in Figure $3$. This suggests
that a free choice at $Q^{\prime }$ could have superluminal consequences at $%
Q$, apparently in conflict with relativity.

Our resolution of this conflict is to note that if $Q$ is outside the
forwards lightcone of $Q^{\prime }$, then in effect we should regard $P$, $%
Q^{\prime }$ \emph{and} $Q$ as on some hypersurface of simultaneity, which,
like $P\cup Q^{\prime }$ and $P^{\prime }\cup Q$, cannot be identified with
a single inertial frame. There is no requirement in relativistic quantum
mechanics to restrict all experiments to single inertial frames. What
matters is quantum causality. We see from this discussion that the principle
of quantum causality has to be applied not just with respect to hyperplanes
of simultaneity in inertial frames, but for arbitrary spacelike
hypersurfaces as well. Quantum causality should hold even in those
experiments where various pieces of apparatus lie in different inertial
frames. Given this observation, we deduce that after the state is prepared
at $S$, an electron can be detected at $P$, a positron detected at $Q$ and
nothing detected at $Q^{\prime }$ for some runs, whilst for other runs, an
electron would be detected at $P$, a positron detected at $Q^{\prime }$, and
nothing detected at $Q$. There would be no possibility of using this
experiment to signal from $Q^{\prime }$ to $Q$, if $Q$ lay outside the
forwards lightcone centred on $Q^{\prime }$.

We denote the Hilbert space involved in this extended experiment by $%
\mathcal{H(}P\cup Q\cup Q^{\prime }\mathcal{)}\equiv \mathcal{H}(P)\otimes
\mathcal{H}\left( Q\cup Q^{\prime }\right) $. We observe that for this
experimental setup, $\mathcal{H(}Q\cup Q^{\prime }\mathcal{)}$ cannot be the
tensor product $\mathcal{H}(Q)\otimes \mathcal{H(}Q^{\prime }\mathcal{)}$,
because any detection of a positron at $Q^{\prime }$ rules out the detection
of a positron at $Q$, and vice-versa. Instead, we would have $\mathcal{H}%
(Q\cup Q^{\prime })=\mathcal{H}(Q)\oplus \mathcal{H(}Q^{\prime })$, i.e., $%
\mathcal{H}\left( Q\right) $ and $\mathcal{H(}Q^{\prime }\mathcal{)}$ can be
regarded as mutual orthogonal complements in $\mathcal{H}(Q\cup Q^{\prime })$%
.  It should be possible to
use standard quantum relativistic field theory to determine the respective
outcome probabilities at $Q$ and $Q^{\prime }$. These will depend on
where the detectors are situated.

Finally, we comment on the meaning of entanglement. It is not physically
meaningful to talk about the preparation of quantum states, entangled or
not, without reference to any context of subsequent observation. It is the
choice of test apparatus which determines whether a state is to be regarded
as entangled or not. For instance, in Figure $3$, we could not locate $S$ on
the quantum horizon $O$ and meaningfully talk about an entangled
electron-positron state from the point of view of any tests involving pieces
of equipment on either side of the quantum horizon. In any discussion
involving quantum information and black hole physics, problems will
inevitably arise whenever quantum states are discussed without due regard for the
equipment used to test them. For this reason, all discussions of quantum mechanics
across event horizons or wavefunctions
for the universe without due reference to observers and their equipment
should be avoided as metaphysical.

\end{document}